\documentstyle[12pt,epsbox]{article}
\renewcommand{\baselinestretch}{1.22}
\begin{document}
%
\begin{flushright}
  August, 1999 \ \ \\
  OU-HEP-324
\end{flushright}
\vspace{0mm}
\begin{center}
\large{Do we expect light flavor sea-quark asymmetry also for \\
the spin-dependent distribution functions of the nucleon ?}
\end{center}
\vspace{0mm}
\begin{center}
M.~Wakamatsu\footnote{Email \ : \ wakamatu@miho.rcnp.osaka-u.ac.jp}
\end{center}
\vspace{-4mm}
\begin{center}
Department of Physics, Faculty of Science, \\
Osaka University, \\
Toyonaka, Osaka 560, JAPAN
\end{center}
\vspace{0mm}
\begin{center}
T.~Watabe\footnote{Email \ : \ watabe@rcnp.osaka-u.ac.jp}
\end{center}
\vspace{-4mm}
\begin{center}
Research Center for Nuclear Physics (RCNP), \\
Osaka University, \\
Ibaraki, Osaka 567, JAPAN
\end{center}

\vspace{8mm}
\ \ \ \ \ \ PACS numbers : 12.39.Fe, 12.39.Ki, 12.38.Lg, 13.40.Em

\vspace{10mm}
\begin{center}
\small{{\bf Abstract}}
\end{center}
\vspace{-1mm}
\begin{center}
\begin{minipage}{15.5cm}
\renewcommand{\baselinestretch}{1.0}
\small
\ \ \ After taking account of the scale dependence by means of
the standard DGLAP evolution equation, the theoretical predictions
of the chiral quark soliton model for the unpolarized and
longitudinally polarized structure functions
of the nucleon are compared with the recent high energy data.
The theory is shown to explain all the qualitative
features of the experiments, including the NMC data for $F_2^p (x) -
F_2^n (x)$, $F_2^n (x) / F_2^p (x)$, the Hermes and NuSea data for
$\bar{d}(x) - \bar{u}(x)$, the EMC and SMC data for $g_1^p(x)$,
$g_1^n(x)$ and $g_1^d(x)$. Among others, flavor asymmetry of
the longitudinally polarized sea-quark distributions is a remarkable
prediction of this model, i.e., it predicts that $\Delta \bar{d}(x)
- \Delta \bar{u}(x) = C \,x^{\alpha} [ \bar{d}(x) - \bar{u}(x)]$
with a sizable negative coefficient $C \simeq -2.0$ (and 
$\alpha \simeq 0.12$) in qualitative consistency with the recent
semi-phenomenological analysis by Morii and Yamanishi.

\normalsize
\end{minipage}
\end{center}
\renewcommand{\baselinestretch}{2.0}

\newpage

Undoubtedly, the most natural and widely-accepted explanation of
the light flavor sea-quark asymmetry in the nucleon,
revealed by the NMC measurements \cite{NMC91},
would be the one by the pion cloud effects
\cite{HM90}--\cite{Kumano98}.
Accepting its essential validity, the NMC observation
may be taken as the first clear evidence manifesting
the nonperturbative QCD dynamics of spontaneous chiral symmetry
breaking in high-energy deep-inelastic observables.
This flavor asymmetry of the
spin-averaged sea-quark distributions could have been predicted
by some theoretical models which properly incorporate the chiral
symmetry, but unfortunately, most theoretical explanations were
given only after the NMC observation.

Now, a natural question is ``Do we expect light flavor sea-quark
asymmetry not only for the unpolarized distributions but also for the
spin-dependent ones?'' In our opinion, to make a trustworthy
prediction for spin-dependent sea-quark distributions,
the chiral symmetry would be a minimum ingredient to be incorporated
into model construction. Otherwise, it would be hard to explain
the already-established flavor asymmetry
of the unpolarized sea-quark distributions. Also desirable is
a reasonable reproduction of the
existing high energy data for the unpolarized and the longitudinally
polarized structure functions.

We claim that the chiral quark soliton model (CQSM) of the nucleon
\cite{DPP88}-- \cite{WK99} fulfills the above requirements and
that it is in this sense qualified as a good candidate which is
able to give some definite prediction for the flavor asymmetry of the
spin-dependent sea-quark distributions.
In the present report, we shall first compare the theoretical
predictions of the CQSM with
the recent high energy data for the unpolarized as well as the
longitudinally polarized structure
functions, including the NMC data for  $F_2^p (x) -
F_2^n (x)$, $F_2^n (x) / F_2^p (x)$ \cite{NMC91},
the HERMES \cite{HERMES} and NuSea (E866) data \cite{E866} for
$\bar{d}(x) - \bar{u}(x)$, the EMC \cite{E143},\cite{E154},\cite{E155}
and SMC data \cite{SMC99} for $g_1^p(x)$,
$g_1^n(x)$ and $g_1^d(x)$. After confirming that the theory
is able to explain all the qualitative feature of these
experimental data, we finally give a prediction for the
longitudinally polarized anti-quark distributions $\Delta \bar{d}(x)
- \Delta \bar{u}(x)$ in the proton. 

The chiral quark soliton model (CQSM) is a very simple model of the
nucleon \cite{DPP88},\cite{WY91}, which maximally takes account
of the spontaneous chiral symmetry breaking of the QCD vacuum.
It is specified by the effective lagrangian :
\begin{eqnarray}
   {\cal L}_0 = \bar{\psi} \,(\,i \! \not\!\partial - 
   M e^{\,i \gamma_5 \mbox{\boldmath $\tau$}
   \cdot \mbox{\boldmath $\pi$} (x) / f_\pi \,}
   \,) \psi , \label{modlag}
\end{eqnarray}
which describes the effective quark fields with a dynamically
generated mass $M$, interacting with massless (or nearly massless)
pions. The theory is not a renormalizable one and it is defined with
some ultraviolet cutoff $\Lambda$. The above effective lagrangian
was originally introduced by Diakonov and Petrov on the basis of the
instanton-liquid picture of the QCD vacuum \cite{DP86}.
In this derivation,
the dynamical quark mass $M$ and the physical cutoff $\Lambda$
can both be estimated as functions of the average instanton size
$\bar{\rho}$ with $1 / \bar{\rho} \simeq 600 \,\mbox{MeV}$ and
the average separation between instantons $\bar{R}$ with
$1 / \bar{R} \simeq 200 \,\mbox{MeV}$. Their estimate gives
$M \simeq 350 \,\mbox{MeV}$ and $\Lambda \simeq 1 / \bar{\rho} \simeq
600 \,\mbox{MeV}$. To be more precise, the predicted dynamical
quark mass $M$ is momentum dependent, but in its application to
the chiral soliton problem, it is customary to approximate it by
a momentum independent constant. At the same time, the ultraviolet
cutoff is fixed by the condition that the effective meson lagrangian
derived from (\ref{modlag}) reproduces the correct normalization
of the pion kinetic term.
For instance, in the Pauli-Villars regularization
scheme, which is used throughout the present analysis,
what plays the role of a ultraviolet cutoff is the Pauli-Villars
mass $M_{PV}$ obeying the relation :
\begin{equation}
   \frac{N_c M^2}{4 \pi^2} \ln {\left( \frac{M_{PV}}{M} \right)}^2 = 
   f_\pi^2 ,
\end{equation}
with $f_\pi$ the pion weak decay constant. Using the value of
$M \simeq 375 \,\mbox{MeV}$, which is favored from the phenomenology
of nucleon low energy observables, this gives $M_{PV} \simeq 562
\,\mbox{MeV}$ in qualitative consistency with the estimate
based on the instanton picture \cite{DPPPW96}.
(We refer to \cite{KWW99} for the justification of the
single-subtraction Pauli-Villars scheme in the studies of
quark distribution functions.)
Since we are to use these values of $M$ and $M_{PV}$, there is
no free parameter additionally introduced in the calculation of
distribution functions. 
The model is also compatible with the idea of large $N_c$ QCD
advocated by Witten many years ago \cite{Witten84}.
According to him, in the limit
of $N_c = \infty$, the QCD is equivalent to an effective theory of
mesons, and a baryon appears as a topological soliton of this
effective meson lagrangian. Though an appealing idea, a practical
problem of this scenario is that the relevant effective meson theory
may not be so simple as that of the Skyrme model. In fact, who can
imagine an effective meson theory, which is able to describe
deep-inelastic scattering observables of the nucleon?
On the contrary, the chiral quark soliton model is an effective
theory, which realizes the idea of large $N_c$ QCD in more economical
way. In fact, in the stationary-phase approximation, which is exact
in the large $N_c$ limit \cite{DPP88}, the theory is known to have
a self-consistent soliton-like solution of hedgehog
shape. The energy degeneracy of this soliton configuration under
the spatial and isospin rotation
induces spontaneous rotational motion of hedgehog soliton.
The semiclassical quantization of this collective rotation
leads to a systematic method of calculation of any nucleon
observables, which is given in a perturbative series in the
collective angular velocity operator $\Omega$.
(This takes the form of a $1 / N_c$ expansion, since $\Omega$ itself
is an $1/N_c$ quantity.) The numerical method of calculation of the
static nucleon observables was 
established many years ago \cite{WY91}, and it has been successfully
applied to many interesting observables \cite{Reviews}.
However, the calculation of quark
distribution functions, especially with full inclusion of the vacuum
polarization contributions, is quite involved. The calculation of all
the combinations of the twist-2 distributions has been completed
only recently \cite{DPPPW96}, \cite{WK98}--\cite{WK99}.

  We first summarize in Fig.1 our final results for the unpolarized
and longitudinally polarized distribution functions of the leading
twist 2. (In the present investigation, some refinement has been
achieved in the numerical method of calculation of these distribution
functions in the small $x$ region. As a consequence, there are some
minor changes from the results reported in our previous
paper \cite{WK99}, especially for the distribution
$\Delta u(x) + \Delta d(x)$. However, these
changes are so small that they hardly leave appreciable
influence on the structure functions obtained after
$Q^2$-evolution.)

\begin{figure}[htbp] \centering
\psbox[width=15cm]{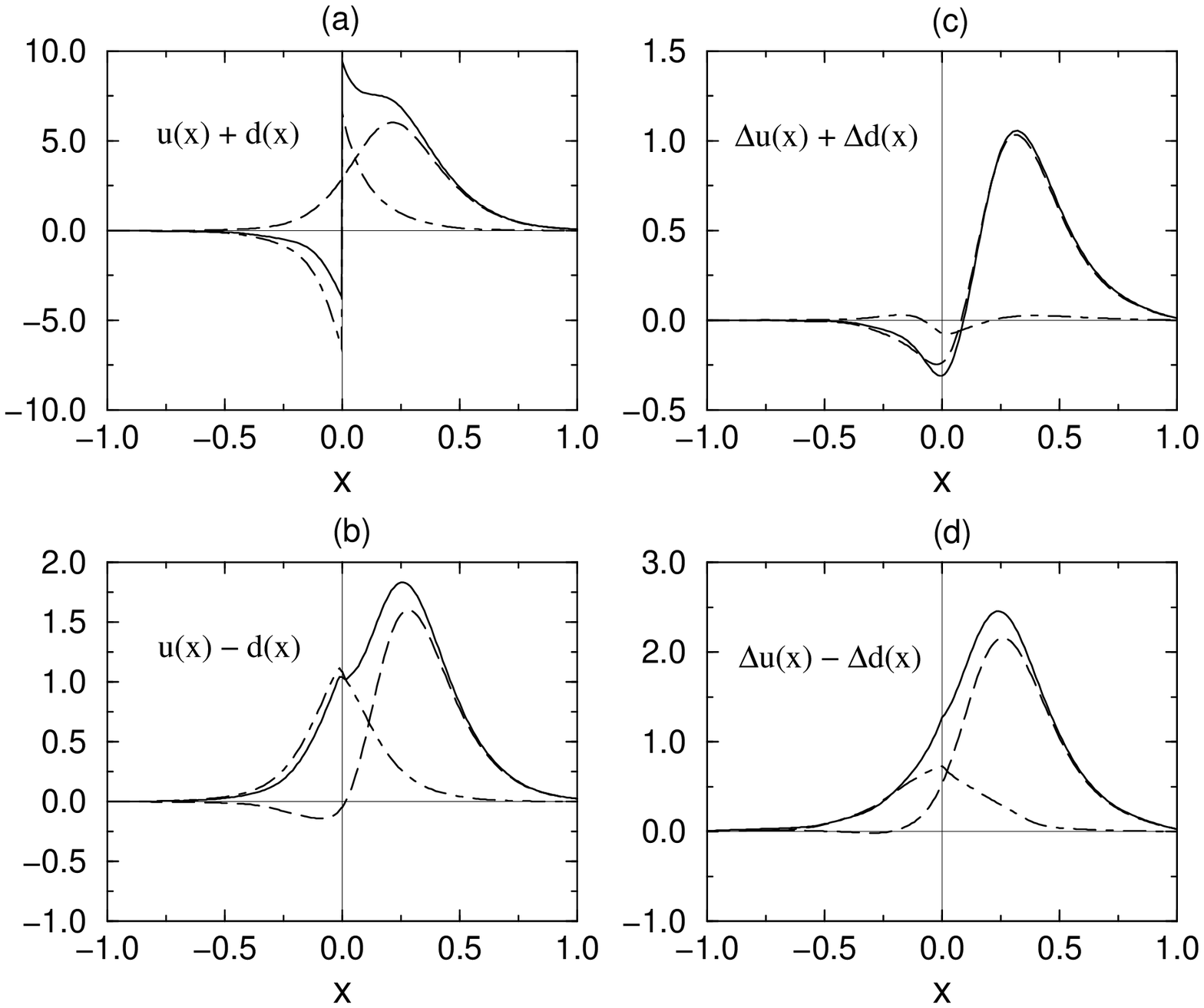}
\renewcommand{\baselinestretch}{1.00}
\caption{The theoretical predictions of the CQSM for the unpolarized
distributions $u(x) + d(x)$ and $u(x) - d(x)$ as well as for the
longitudinally polarized distributions $\Delta u(x) + \Delta d(x)$
and $\Delta u(x) - \Delta d(x)$. In all the figures, the long-dashed
and dash-dotted curves respectively stand for the contributions of
the discrete valence level and that of the Dirac continuum in the
self-consistent hedgehog background, whereas their sums are shown
by the solid curves. The distribution functions in the negative
$x$ region are to be interpreted as the antiquark ones according to
the rules (\ref{antiq1})--(\ref{antiq4}).}
\renewcommand{\baselinestretch}{1.32}
\end{figure}

We recall that, within the theoretical framework of the flavor
SU(2) CQSM, any nucleon observables are evaluated in either of
the isoscalar combination or the isovector one \cite{DPP88},\cite{WY91}.
In fact, theoretical formulas
for them have totally dissimilar form reflecting the fact that they
have different dependence on the collective angular
velocity $\Omega$ given as \cite{DPPPW96},\cite{PPGWW99},\cite{WK99}
\begin{eqnarray}
  u(x) + d(x) \ &\sim& \ N_c \,O (\Omega^0) \ \sim \ O (N_c^1), \\
  u(x) - d(x) \ &\sim& \ N_c \,O (\Omega^1) \ \sim \ O (N_c^0), \\
  \Delta u(x) + \Delta d(x) \ &\sim& \  
  N_c \,O (\Omega^1) \ \sim \ O (N_c^0), \\
  \Delta u(x) - \Delta d(x) \ &\sim& \ N_c \,
  [ \,O (\Omega^0) + O (\Omega^1) \,]
  \ \sim \ O (N_c^1) + O(N_c^0) ,
\end{eqnarray} 
where use has been made of the fact that $\Omega$ scales as $1 / N_c$.
In Fig.1, quark distributions with negative $x$ are to be interpreted
as antiquark distributions according to the rule \cite{DPPPW96} :
\begin{eqnarray}
  u (-x) + d (-x) &=& -[\bar{u} (x) + \bar{d} (x)],
  \hspace{12mm} (0<x<1), \label{antiq1} \\
  u (-x) - d (-x) &=& -[\bar{u} (x) - \bar{d} (x)],
  \hspace{12mm} (0<x<1), \label{antiq2} \\
  \Delta u (-x) + \Delta d (-x) &=& 
  \Delta \bar{u} (x) + \Delta \bar{d} (x),
  \hspace{11mm} (0<x<1), \label{antiq3} \\
  \Delta u (-x) - \Delta d (-x) &=& 
  \Delta \bar{u} (x) - \Delta \bar{d} (x), 
  \hspace{11mm} (0<x<1). \label{antiq4}
\end{eqnarray}
Here, the long-dashed and dash-dotted curves respectively
stand for the contribution of the discrete valence level
(it is a deep bound state emerging from the positive energy continuum
under the influence of hedgehog mean field) and that of the
negative energy Dirac sea (the latter is sometimes called the vacuum
polarization contribution), while their sums are shown by the
solid curves. We emphasize that the separation into the
valence and sea-quark contributions is highly model-dependent
concept, and too much significance should not be given to it.
Let us first look into $u (x) + d (x)$ in Fig.1(a). This distribution,
which emerges at the leading order of $N_c$, was first evaluated by
Diakonov et al. \cite{DPPPW96}. 
As emphasized by them, the proper account of the
vacuum polarization contribution is essential here. 
In fact, the ``valence-quark-only'' approximation would lead to
$\bar{u}(x) + \bar{d}(x) < 0$ \cite{WGR96},
thereby violating the positivity
of the parton distribution. After including the vacuum polarization
contribution, on the other hand, this fundamental requirement
is satisfied finely.
Next, Fig.1(b) shows the isovector combination of the
unpolarized distribution function $u(x) - d(x)$. The vacuum
polarization contribution is sizable also for this quantity.
Among others, the significant positivity of $u(x) - d(x)$ in the
negative $x$ region denotes that $\bar{u}(x) - \bar{d}(x) < 0$ for
physical value of $x$ with $0 < x < 1$, thereby indicating the
excess of $\bar{d}$ sea over the $\bar{u}$ sea in the proton.
This result follows from
the fact that the effect of pion cloud is automatically incorporated
into the framework of the CQSM \cite{Wakam93}.
We shall later compare the above
prediction of the CQSM directly with the existing high energy data.

Next, we turn to the longitudinally polarized distributions.
The isoscalar combination $\Delta u(x) + \Delta d(x)$ and the
isovector one $\Delta u(x) - \Delta d(x)$ are respectively shown
in Fig.1(c) and Fig.1(d). The isoscalar distribution
survives only at the first order of $\Omega$ or
equivalently at $O(N_c^0)$. This is because the leading-order term
vanishes identically because of the hedgehog symmetry.
On the other hand, the leading contribution to
the isovector distribution arises from the $O(\Omega^0)$ or $O(N_c^1)$
term. Although the main feature of the isovector longitudinally
polarized distributions are controlled by this leading order
contribution, the next-to-leading-order contribution is also
important. In fact, without this $1 / N_c$ correction, the
first moment of $\Delta u(x) - \Delta d(x)$, i.e. the isovector
axial charge of the nucleon $g_A^{(3)}$ would be largely
underestimated. (Incidentally, the presence of this first order
rotational correction to $g_A^{(3)}$ is a distinguishable feature
of our effective quark theory as compared with a deeply connected
soliton model of the nucleon, i.e. the Skyrme model
\cite{WW93},\cite{CBGPPWW94}. This breakdown of the
so-called Cheshire-cat principle can be understood as originating
from the noncommutativity of the bosonization procedure and the
collective quantization procedure of the rotational
motion \cite{Wakam96}.)
The numerical results also shows that the isoscalar and isovector
distributions have rather different behaviors as functions of $x$.
The first thing one may notice for the isoscalar distribution
is the smallness of the sea quark contribution. (At least the
smallness of the sea quark contribution to the first moment
of $\Delta u(x) + \Delta d(x)$ has long been known to us \cite{WY91}.)
The smallness of the sea quark contribution here does not necessarily
mean negligible role of the vacuum polarization effect itself.
Since our valence quark level is a deep bound state generated by
the strong mean-field of hedgehog shape, it is legitimate to think
that the valence level contribution above also contains some sort of
vacuum polarization effect due to this strong mean field.
In fact, the first moment of $\Delta u(x) + \Delta d(x)$, which
can be interpreted as the quark spin content of the nucleon
$\Delta \Sigma_3$ or the flavor-singlet axial charge at the
energy scale of the present model, turns out to be fairly small as
\begin{eqnarray}
   \Delta \Sigma_3 &\equiv& \int_{-1}^{1} dx \,[ \,\Delta u(x) +
   \Delta d(x) \,] \nonumber \\
   &=& \int_0^1 dx \,[ \,\Delta u(x) + \Delta \bar{u}(x) +
   \Delta d(x) + \Delta \bar{d}(x) \,]
   \ \simeq \ 0.35 . 
\end{eqnarray}
As advocated in \cite{WY91}, the symmetry breaking strong mean-field
of hedgehog shape must be responsible for this smallness of the quark
spin content. A quite interesting feature of the theoretical
distribution $\Delta u(x) + \Delta d(x)$ is that it changes sign
from positive to negative as $x$ becomes smaller than about 0.1.
As already pointed out in our previous paper \cite{WK99},
what causes this
sign change is the first order rotational correction, which arises
from the proper account of the nonlocality (in time) of the
quark bilinear operator entering into the theoretical definition
of the quark distribution function. We shall later show that
this sign change of the isoscalar distribution function
in the small $x$ region is just what is required
by the recent experimental data for the deuteron structure
function $g_1^d (x)$. 

The isovector combination $\Delta u(x) - \Delta d(x)$ illustrated
in Fig.1(d) shows totally different $x$ dependence from the
isoscalar one. The vacuum polarization contribution to this
distribution function has a sizable magnitude of peak with
positive sign around $x \simeq 0$. A sizable nonzero value of
$\Delta u(x) - \Delta d(x)$ in the negative $x$ domain has a
far-reaching physical consequence. That is, it means the violation of
SU(2) symmetry for the longitudinally polarized sea-quark
(anti-quark) distributions. In consideration of the relation
(\ref{antiq4}), we can read from Fig.1(d) that
$\Delta \bar{u}(x) - \Delta \bar{d}(x) > 0$ for physical range of
$x$ between 0 and 1.

To sum up, within the theoretical framework of the CQSM, the
isoscalar and isovector distribution functions turn out to
have fairly different $x$ dependence reflecting the fact that
they have totally dissimilar $N_c$ dependence. Now we shall
compare these predictions of the CQSM with the existing high
energy data for the unpolarized as well as the
longitudinally polarized structure functions.
Before carrying out this comparison, we must first take account
of the scale dependence of the distribution functions. We have
done it by using the Fortran programs given by
Saga group \cite{HKM98}, which
provide us with solutions of
Dokshitzer-Gribov-Lipatov-Altarrelli-Parisi
(DGLAP) evolution equation at the next-to-leading order (NLO).
The question here is what value we should use for the initial
energy scale of this $Q^2$ evolution. Here we have tried to see
the effect of variation of $Q^2_{init}$ in a range centered
around the model energy scale of $M^2_{PV} \simeq 
{(0.56 \,\mbox{GeV})}^2 \simeq 0.31 \,\mbox{GeV}^2$. We find
that the final results are not significantly different in the
range $0.25 \mbox{GeV}^2 \leq Q^2_{init} \leq 0.35 GeV^2$.
We shall use the value $Q^2_{init} = 0.30 \,\mbox{GeV}^2$
throughout the following analyses. In solving the
DGLAP equation, we set $N_f = 3$ and assume $s(x) = \bar{s}(x) = 0$,
$g(x) = 0$ and $\Delta s(x) = \Delta \bar{s}(x) = 0, \Delta g(x) = 0$
at the initial energy scale.

\begin{figure}[htbp] \centering
\psbox[width=10cm]{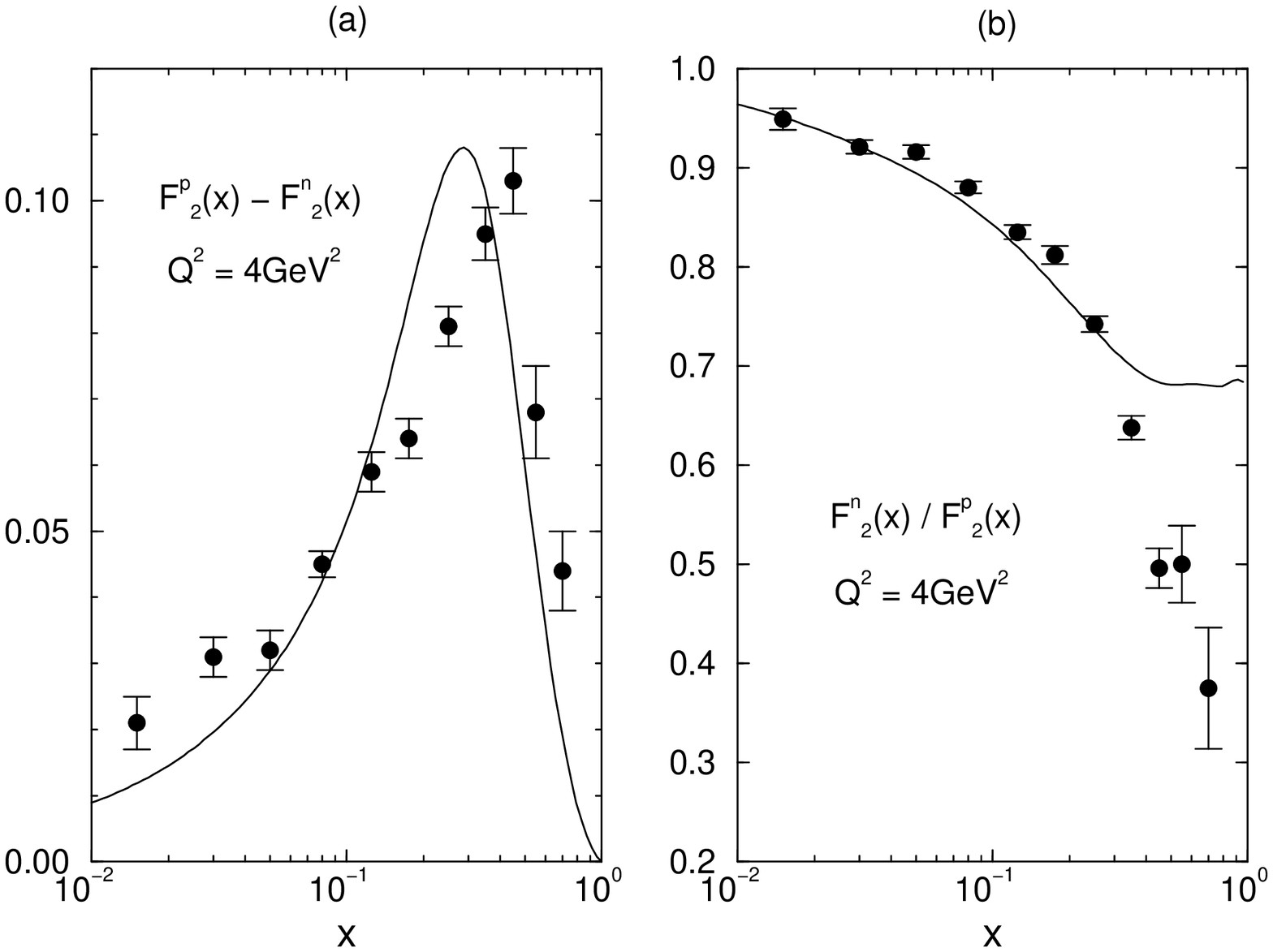}
\renewcommand{\baselinestretch}{1.00}
\caption{The predictions for $F^p_2 (x) - F^n_2 (x)$ and
$F^n_2 (x) / F^p_2 (x)$ at $Q^2 = 4 \,GeV^2$ are compared with the
NMC data given at the corresponding energy scale \cite{NMC91}.}
\renewcommand{\baselinestretch}{1.32}
\end{figure}
\vspace{4mm}
\noindent

Fig.2 shows the theoretical predictions for $F^p_2 (x) - F^2_n (x)$
and $F^n (x) / F^p_2 (x)$ at $Q^2 = 4 \,\mbox{GeV}^2$ in comparison
with the NMC data. One sees that the NMC data for
$F^p_2 (x) - F^2_n (x)$ are well reproduced by the theory, except
for a small discrepancy between the peak positions of the
theoretical curve and the experimental data. Next, we compare our
theoretical prediction for the ratio $F^n_2 (x) / F^p_2 (x)$
with the corresponding NMC data. One can say that the agreement
between the theory and the experiment is fairly good at least
for $x \leq 0.25$.
The theoretical prediction
deviates from the data as $x$ becomes larger than $0.3$. This might be
connected with the fact that our treatment of the soliton
center-of-mass motion is essentially nonrelativistic.
Note, however, that a reliable theoretical prediction of the ratio
$F^n (x) / F^p (x)$ near $x \simeq 1$ is extremely difficult,
because both of $F^n (x)$ and $F^p (x)$ damps rapidly
as $x$ becomes larger.

\vspace{2mm}
\begin{figure}[htbp] \centering
\psbox[width=10cm]{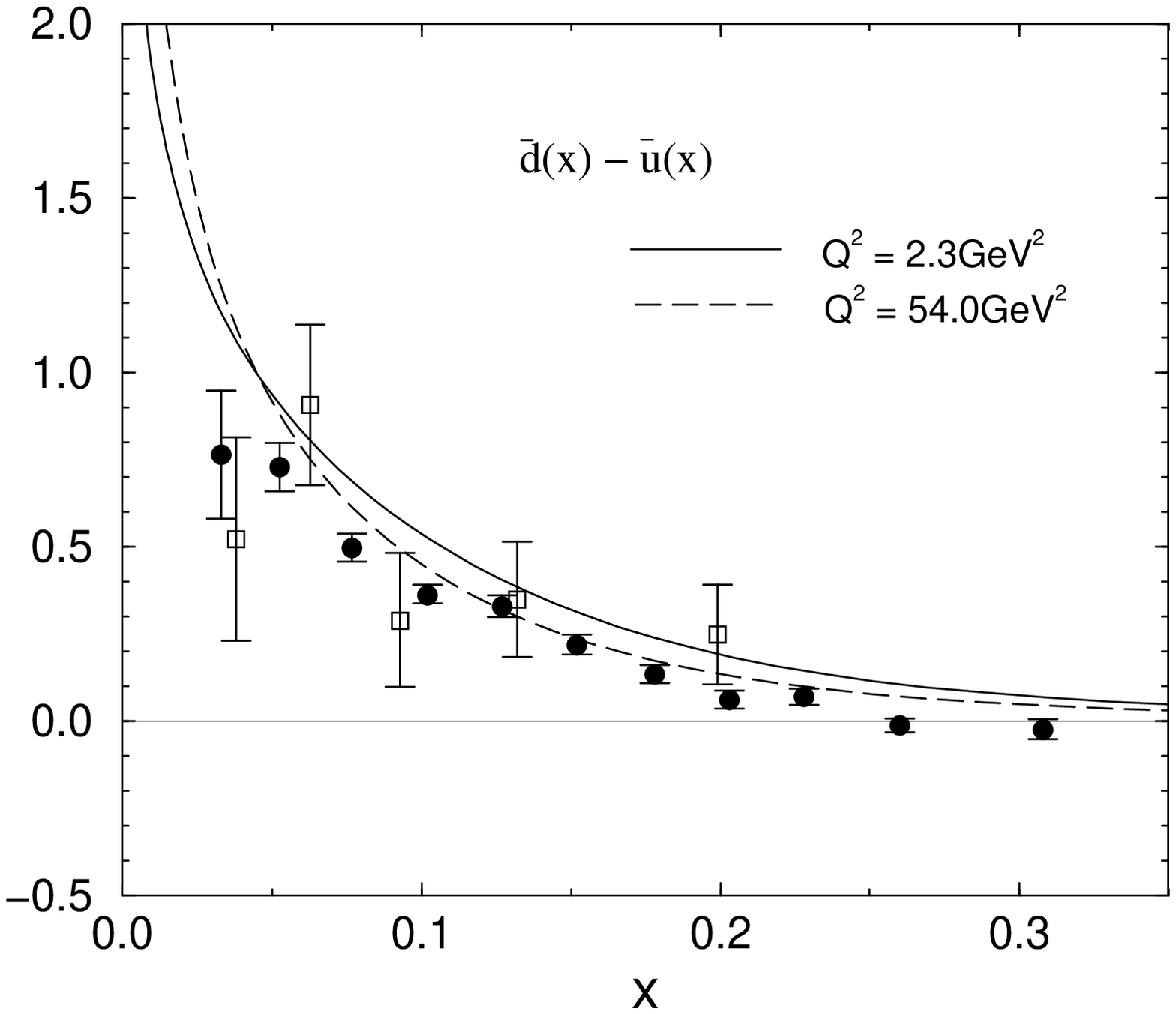}
\renewcommand{\baselinestretch}{1.00}
\caption{The theoretical predictions for the unpolarized antiquark
distribution $\bar{d}(x) - \bar{u}(x)$ at $Q^2 = 2.3 \,\mbox{GeV}$
and $Q^2 = 54 \,\mbox{GeV}$ in comparison with the 
HERMES \cite{HERMES} and the E866 data \cite{E866}.
Here the open squares represent the HERMES data given at
$Q^2 = 2.3 \,\mbox{GeV}$, while filled squares denote the
E866 data given at $Q^2 = 54 \,\mbox{GeV}$.}
\renewcommand{\baselinestretch}{1.32}
\end{figure}

Next, in Fig.3, we compare the theoretical
predictions for the anti-quark distribution $\bar{d}(x) - \bar{u}(x)$
evaluated at two different values of $Q^2$ with the HERMES
\cite{HERMES} and NuSea data \cite{E866} given at the
corresponding energy scales.
(One certainly confirms that the scale dependence of the quantity
$\bar{d}(x) - \bar{u}(x)$ is rather small.)
The general trend of the experimental data is well reproduced
by the theory, although the theory has a tendency
to slightly overestimate the magnitude of the flavor asymmetry
of the sea-quark distributions.   

Now we turn to the discussion of the longitudinally polarized
distributions. Shown in Fig.4 are the longitudinally polarized
structure functions for the proton, the neutron and the deuteron,
$g^p_1 (x), g^n_1 (x)$ and $g^d_1 (x)$.
The results for $g^p_1 (x)$ and $g^n_1 (x)$ were already reported
in \cite{WK99}, but we show them again for the purpose of comparison,
since $g^d_1 (x)$ is essentially given as their average.
To be more precise, we evaluate $g^d_1 (x,Q^2)$ in Fig.4(c)
by making use of the standardly-used formula
\begin{equation}
   g^d_1 (x,Q^2) = \frac{1}{2} \,( \, 
   g^p_1 (x,Q^2) + g^n_1 (x,Q^2) \,) \,( \, 1 - 1.5 \omega_D \,) ,
\end{equation}
with $\omega_D$ the probability that the deuteron is in a D-state.

\begin{figure}[htbp] \centering
\psbox[width=15cm]{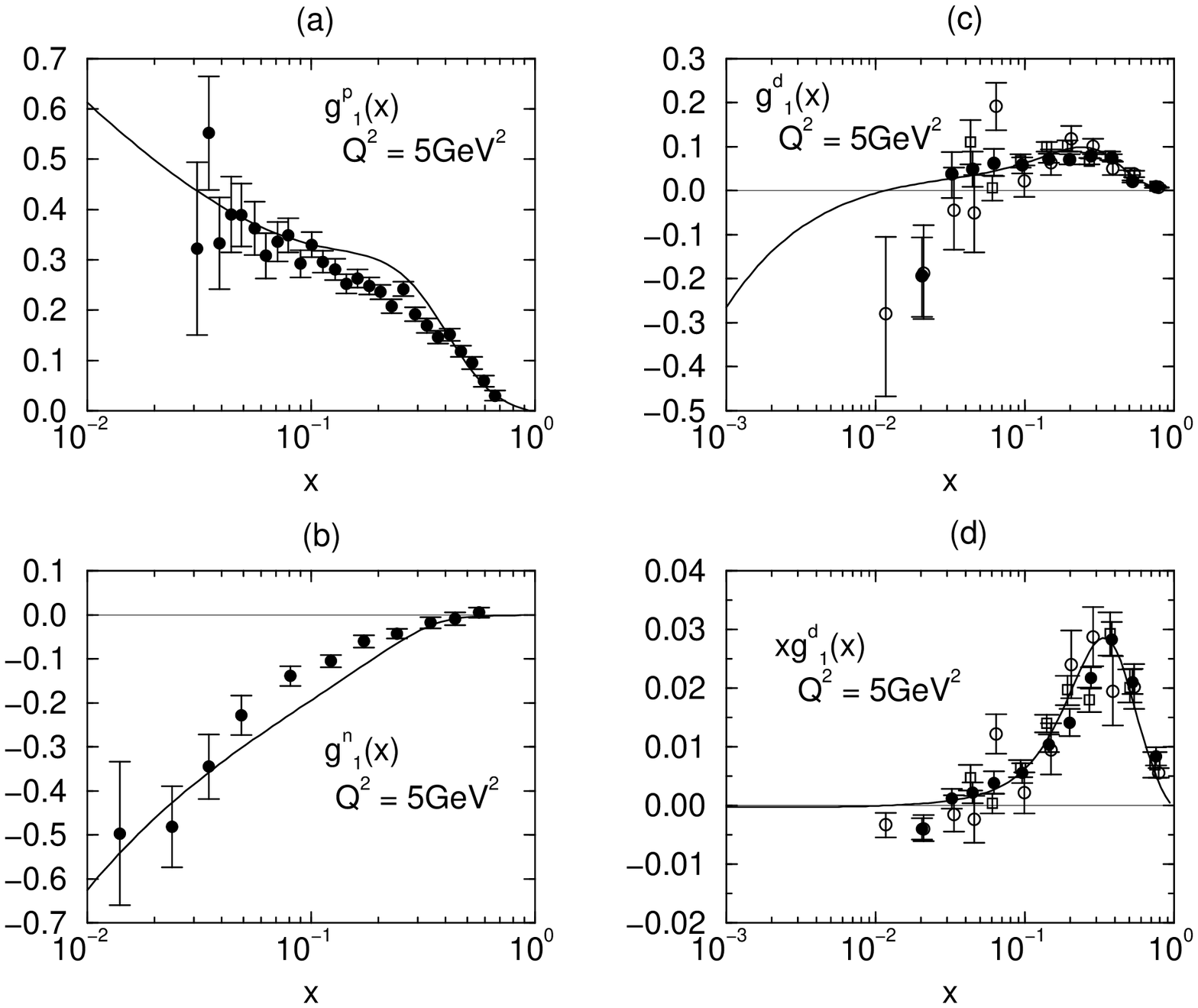}
\renewcommand{\baselinestretch}{1.00}
\caption{The theoretical predictions for the longitudinally polarized
structure functions for the proton, the neutron and the deuteron
at $Q^2 = 5 \,\mbox{GeV}^2$ in comparison with the corresponding
experimental data. The filled circles in (a) and (b) respectively
corresponds to the E143 \cite{E143} and the E154 data \cite{E154},
whereas the filled circles, the open circles and the open squares
in (c) and (d) represent the E143, the E155 \cite{E155} and
the SMC data \cite{SMC99}.}
\renewcommand{\baselinestretch}{1.32}
\end{figure}

As already argued in \cite{WK99}, a prominent feature of
the CQSM is the
good reproduction of the neutron data, which are known to have large
magnitudes with negative sign. We have interpreted this as a
manifestation of the chiral symmetry, maximally incorporated
into this effective quark model. Now a direct comparison with the
deuteron data reveals another (though related) interesting aspect
of physics. An noticeable characteristic of the recent data for
$g^d_1 (x,Q^2)$ is that it appears to show sign change as $x$
approaches zero, although care must be paid to the fact that
the precision of the experimental data in the small $x$ region
is still poor. Quite interestingly, the theoretical prediction
of the CQSM closely follows the variation of this $g^d_1 (x,Q^2)$
at least qualitatively. What is the origin of this behavior of the
theoretical structure function? Undoubtedly, it can be traced back
to the behavior of the isoscalar distribution function
$\Delta u(x) + \Delta d(x)$ shown in Fig.1(c).
In fact, as emphasized by Windmolders \cite{Windmol99},
the singlet distribution is
strongly constrained by the measured value of $g^d_1$. (This seems
only natural, since the deuteron is an isoscalar target.)
To see it more closely, we recall the general expression for $g^d_1$
given as

\begin{equation}
  g^d_1 \ \sim \ \frac{1}{9} \,\left( \,
  \frac{1}{4} C_{NS} \otimes \Delta q_8 + 
  C_S \otimes \Delta \Sigma + 
  2 N_f C_g \otimes \Delta g \,\right) ,
\end{equation}
where
\begin{eqnarray}
  \Delta q_8 &=& \Delta u + \Delta d - 2 \Delta s , \\
  \Delta \Sigma &=& \Delta u + \Delta d + \Delta s ,
\end{eqnarray}
and $C_{NS}, C_S, C_G$ are the non-singlet, singlet and gluon
Wilson coefficients and the symbol $\otimes$ represents
convolution with respect to $x$.

Assuming minor role of the $s$-quark degrees of freedom at the
energy scale in question, the difference between $\Delta q_8$
and $\Delta \Sigma$ would be small. The $g^d_1$ is then determined
by $\Delta u + \Delta d$ and $\Delta g$ at this energy scale.
Now, somewhat peculiar small-$x$ behavior of the isoscalar
longitudinally polarized distribution function predicted by the
CQSM seems to get a phenomenological support of the latest
deuteron data.
In any case, the $g^d_1$, especially in a smaller $x$ region, appears
to be extremely sensitive to the detailed dynamical content of the
theories. To be able to reproduce it or not would then be a good
touchstone of model selection.

\vspace{2mm}
\begin{figure}[htbp] \centering
\psbox[width=10cm]{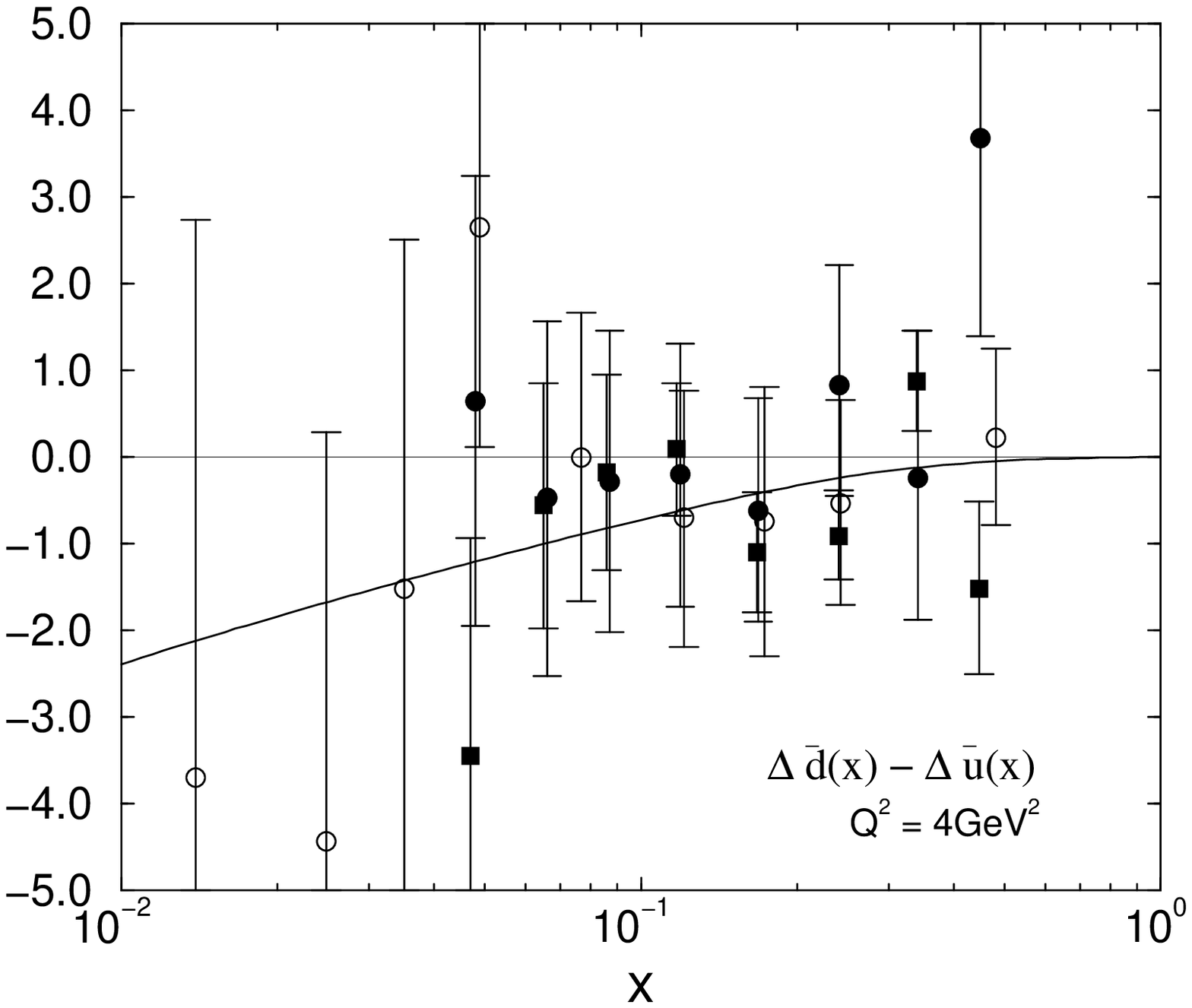}
\renewcommand{\baselinestretch}{1.00}
\caption{The theoretical prediction for the longitudinally polarized
antiquark distribution $\Delta \bar{d}(x) - \Delta \bar{u}(x)$
at $Q^2 = 4 \,\mbox{GeV}^2$ is compared with the recent
semi-phenomenological fit by Morii and Yamanishi.
See figure caption of \cite{MY99} for the detailed meaning of three
different marks in their results.}
\renewcommand{\baselinestretch}{1.32}
\end{figure}

Now that we have confirmed that the CQSM can explain the principle
experimental data for the unpolarized and the longitudinally
polarized structure functions of the nucleon as well as the deuteron,
we come back to an interesting prediction of the CQSM given by
Fig.1(d), i.e. the possible violation of the light flavor sea-quark
asymmetry for the longitudinally polarized distributions.
Since the sea-quark distributions are not well constrained by the
inclusive data alone, very little is known as to them. 
In a recent
report \cite{MY99}, Morii and Yamanishi proposed new
formulas for extracting
a difference, $\Delta \bar{d} - \Delta \bar{u}$, from data of
the polarized semi-inclusive processes $\vec{l} + \vec{N} 
\rightarrow  l^\prime + H + X$ and estimated the value of it
from the available data of the SMC and Hermes groups.
The results of their analyses are shown in Fig.5 together with the
prediction of the CQSM at the corresponding energy scale,
$Q^2 = 4 \,\mbox{GeV}^2$.
Unfortunately, the uncertainties of the
fit are so large that nothing definite can be said at the present
status. Note, however, that Morii and Yamanishi parameterized
$\Delta \bar{d}(x) - \Delta \bar{u}(x)$ as
\begin{equation}
  \Delta \bar{d}(x) - \Delta \bar{u}(x) = C \,x^\alpha \,
  \left( \,\bar{d}(x) - \bar{u}(x) \,\right)
\end{equation}
and determined the value of $C$ and $\alpha$ from the $\chi^2$-fit
of the results presented in the above figure. The results were
$C = - 1.00$ and $\alpha = 0.18$, thereby indicating an asymmetry of
$\bar{d}$ and $\bar{u}$. The negative value of $C$ denotes that
\begin{equation}
  \Delta \bar{d}(x) - \Delta \bar{u}(x) < 0 ,
\end{equation}
since we already know that $\bar{d}(x) - \bar{u}(x) > 0$.
This is qualitatively consistent with the prediction of the CQSM.
In fact, the values of $C$ and $\alpha$ determined from our
theoretical distributions at $Q^2 = 4 \,\mbox{GeV}^2$ turn out to be
\begin{equation}
  C \simeq - 2.0, \ \ \ \alpha \simeq 0.12 ,
\end{equation}
in qualitative consistency with the result of their analysis.

Despite encouraging phenomenological support to the
present approach, an important question remains to be answered.
The question concerns the role of gluons at the relatively
low energy scale of $Q^2 = 0.3 \sim 0.5 \,\mbox{GeV}^2$, which may
be taken as a starting energy of the DGLAP evolution
equation. We have simply assumed zero for the input gluon densities
at the initial energy scale. This assumption may not be
necessarily justified, especially for the unpolarized case.
In fact, there is some phenomenological indication that the
gluon saturates nearly $30 \,\%$ of the nucleon momentum sum
rule even at such low energy scale \cite {GRV95}.
(Note, however, that the
isovector-type quantities like $F^p_2 (x) - F^n_2 (x)$ and
$\bar{d}(x) - \bar{u}(x)$ investigated in the present paper
would be insensitive to these input gluon densities.)
On the other hand, very little is known for the polarized
gluon densities. A relatively good agreement between the
experimental structure functions and our theoretical
predictions, which are obtained by assuming zero gluon polarization
at $Q^2_{init} = 0.3 \,\mbox{GeV}^2$, indicates that the role of
gluons may be less important than the case of the unpolarized
distributions. (This never denies the importance of gluon
polarization at high energy scale, since the gluons are known to
acquire polarization rapidly through the processes of perturbative
evolution.) Undoubtedly, it is a very important theoretical
question to reliably predict the gluon densities to be used as
input distributions of the renormalization-group evolution.
The importance of the gluon densities at this matching energy scale
of $Q^2 = 0.3 \sim 0.5 \,\mbox{GeV}^2$ was also emphasized by
by Lampe and Reya in a recent review \cite{LR98}.
According to these authors, if low energy models cannot provide
necessary input gluon densities at this energy, they would only refer
to some nonperturbative input quark scale which cannot be
reached by perturbative evolution.

In summary, the CQSM allows us for the calculation of the full
$x$-dependence of the nucleon structure functions without
introducing any new free parameters. In particular, what makes
this approach quite promising is the fact that it predicts,
besides the valence-like distributions, the sea-quark-like
densities concentrated in the small $x$ domain, if it is
treated in a theoretically consistent manner (i.e., not in the
``valence-quark-only'' approximation). This unique feature of the
model is expected to provide us with a good starting point
for theoretically understanding unpolarized data
as well as yet-to-be-obtained polarized data in the small $x$ region.
Making full use of this advantage of the CQSM, we have made a
prediction for the flavor asymmetry of the longitudinally
polarized antiquark distributions in the proton. The model has
been shown to predict $\Delta \bar{d}(x) - \Delta \bar{u}(x) < 0$,
while at the same time $\bar{d}(x) - \bar{u}(x) > 0$, i.e.,
it definitely predicts light flavor sea-quark asymmetry not only
for the spin-averaged distributions but also for the longitudinally
polarized distributions. We hope that this interesting prediction
of the CQSM will be tested through accumulation and analyses of more
precise experimental data on the polarized semi-inclusive processes
in the near future.

\vspace{10mm}
\noindent
\begin{large}
{\bf Acknowledgement}
\end{large}
\vspace{3mm}

One of the author (M.W.) would like to express his sincere thanks to
Prof. M.~Wakai for many helpful discussions.

%
%
\renewcommand{\baselinestretch}{1.16}

\end{document}